# Textual interpretation of transient image classifications from large language models

Fiorenzo Stoppa ®[1,7] ✉, Turan Bulmus[2,7], Steven Bloemen[3], Stephen J. Smartt[1,4], Paul J. Groot ®[3,5,6], Paul Vreeswijk[3] & Ken W. Smith[1,4]

Modern astronomical surveys deliver immense volumes of transient detections, yet distinguishing real astrophysical signals (for example, explosive events) from bogus imaging artefacts remains a challenge. Convolutional neural networks are effectively used for real versus bogus classification; however, their reliance on opaque latent representations hinders interpretability. Here we show that large language models (LLMs) can approach the performance level of a convolutional neural network on three optical transient survey datasets (Pan-STARRS, MeerLICHT and ATLAS) while simultaneously producing direct, human-readable descriptions for every candidate. Using only 15 examples and concise instructions, Google's LLM, Gemini, achieves a 93% average accuracy across datasets that span a range of resolution and pixel scales. We also show that a second LLM can assess the coherence of the output of the first model, enabling iterative refinement by identifying problematic cases. This framework allows users to define the desired classification behaviour through natural language and examples, bypassing traditional training pipelines. Furthermore, by generating textual descriptions of observed features, LLMs enable users to query classifications as if navigating an annotated catalogue, rather than deciphering abstract latent spaces. As next-generation telescopes and surveys further increase the amount of data available, LLM-based classification could help bridge the gap between automated detection and transparent, human-level understanding.

Modern time-domain astronomy hinges on the ability to identify and classify a vast array of ephemeral phenomena, from explosive supernovae and gravitational wave counterparts to flickering variable stars and subtle lensing events, all of which provide crucial insights into the dynamic processes of the Universe. When sources are discovered through high-energy, neutrino or gravitational wave emission, optical identification is always essential for identifying the source with high spatial precision and for providing a redshift, distance and an energy scale (for example, refs. 1–4). Wide-field optical time-domain surveys commonly employ a technique called difference imaging to identify flux changes between a newly taken image and a reference image[5–7]. Such methods produce a flood of transient candidates, many of which are spurious artefacts and generally referred to as 'bogus'. Convolutional neural networks (CNNs) trained on large samples of labelled data have been successful in real versus bogus classification when applied to the image pixels[8–16]. Although CNNs commonly exceed 98% accuracy, their latent representations remain opaque. Understanding why a CNN labels a candidate as real or bogus often demands post hoc methods like saliency maps, which can be limited in their interpretative power[17–19]. The application of CNNs often results in the production of a single number on which the real versus bogus classification is based.

[1]Astrophysics Sub-Department, Department of Physics, University of Oxford, Oxford, UK. [2]Google Cloud, Amsterdam, The Netherlands. [3]Department of Astrophysics/IMAPP, Radboud University, Nijmegen, The Netherlands. [4]Astrophysics Research Centre, School of Mathematics and Physics, Queen's University Belfast, Belfast, UK. [5]Department of Astronomy and Interuniversity Institute for Data Intensive Astronomy, University of Cape Town, Rondebosch, South Africa. [6]South African Astronomical Observatory, Observatory, South Africa. [7]These authors contributed equally: Fiorenzo Stoppa, Turan Bulmus. ✉e-mail: fiorenzo.stoppa@physics.ox.ac.uk

The emergence of large language models (LLMs)[20–23] provides an alternative approach to transient classification. Unlike CNNs, LLMs can generate textual descriptions alongside their classification outputs, facilitating a form of transparency that aligns with human reasoning. To evaluate their effectiveness, we applied Google's LLM, Gemini, to data collected from three main wide-field time-domain surveys: Pan-STARRS (1.8-m telescope), MeerLICHT (0.65 m) and ATLAS (0.5 m). The MeerLICHT dataset[24,25] comprises approximately 3,200 candidates with a mixture of explosive transients, variable stars and artefacts. From the ATLAS survey data[26] we selected a set of 2,000 explosive transients and spurious detections[27]. The Pan-STARRS dataset (from the PS1 telescope[28]) similarly includes 2,000 candidates but features a unique type of artefact—chip gaps—caused by the camera design. The Gigapixel Camera 1 on Pan-STARRS has 60 CCDs, each composed of an 8 × 8 subcell array, so chip gaps are prevalent[28], whereas both MeerLICHT and ATLAS are single monolithic silicon devices. Ground-truth labels for these datasets were obtained as follows. For the MeerLICHT dataset, labels were manually assigned by expert astronomers[11]; for ATLAS and Pan-STARRS, approximately 50% of the examples (the real transients) were manually classified, whereas the remaining examples were taken from the garbage lists generated by the respective telescope pipelines. The diversity in the detector pixel scales, spatial resolution and the point-spread-function profile across these datasets ensures a comprehensive assessment of the adaptability of the LLM.

Unlike traditional machine learning models that require hundreds of thousands of labelled examples for training, we employed a few-shot learning approach, providing Gemini with only 15 annotated triplets per telescope. Each triplet consisted of a target image, a reference image and a difference image, and these were accompanied by descriptive text explaining these images that was written by expert astronomers. The 15 examples were manually selected by astronomers to reflect a representative mix of real and bogus cases and to cover a range of typical morphological features relevant to the classification task. In addition, the model was given a simple set of instructions outlining its task. These triplets and their descriptions serve solely as examples to guide the model in adhering to a specific format for its output, which can be adjusted based on the user's goals. The model is not trained on these images but uses them as reference examples to structure its explanations for new inputs. The complete set of MeerLICHT few-shot examples, including images and explanations, is shown in 'Few-shot learning examples for the MeerLICHT dataset'.

Even with this minimal set of images and contextual information, Gemini achieved an average accuracy of 93% across the three surveys, as validated against ground-truth labels available for the three datasets (Table 1). Crucially, the Gemini outputs are not restricted to binary labels: the model can generate for each candidate natural language descriptions of its morphology, brightness changes and possible astrophysical nature.

Figure 1 illustrates the classification process used by Gemini. It displays examples from the MeerLICHT dataset of transient image input (new, reference and difference) alongside the corresponding class labels, textual explanations and follow-up interest scores. In this specific set-up, the interest scoring system is designed to simulate the need for automated prioritization regarding the following up of explosive transients. Explosive events are marked as 'high interest', variable stars as 'low interest' and artefacts as 'no interest'. By providing clear textual descriptions—such as whether the candidate seems to be hosted by a galaxy, isolated (orphan) or situated in a crowded stellar field—the LLM can inform targeted follow-up strategies by helping observers prioritize events that may be astrophysically intriguing in resource-limited and time-sensitive observational environments.

**Table 1 | Performance metrics for Gemini across the three surveys**

| Telescope | Accuracy (%) | Precision (%) | Recall (%) |
|---|---|---|---|
| ATLAS | 91.9 | 88.5 | 94.5 |
| MeerLICHT | 93.4 | 87.7 | 98.7 |
| Pan-STARRS | 94.1 | 95.4 | 93.1 |

## Results

To evaluate the performance and adaptability of Gemini, we tested it on transient data from three surveys: Pan-STARRS, MeerLICHT and ATLAS, thus providing it with various resolutions, pixel sampling, imaging conditions and transient types. Figure 2 shows examples of typical input triplets (new, reference and difference images) from all three surveys. These samples illustrate the variety of observational conditions and source characteristics that Gemini must handle, demonstrating the ability of the model to operate effectively across heterogeneous datasets.

Gemini operated in a few-shot learning regime, using only 15 annotated examples per survey and a concise set of instructions. Despite these minimal input resources, it achieved strong performance across all three surveys, as shown in Table 1. The accuracy metric, calculated as (TP + TN)/(TP + TN + FP + FN), measures the proportion of correctly classified examples. Precision, defined as TP/(TP + FP), evaluates the fraction of positive classifications that are correct, whereas recall, TP/(TP + FN), reflects the ability to identify all relevant instances. Here, TP stands for true positive, TN for true negative, FP for false positive and FN for false negative. Gemini demonstrated precision and recall values on par with those of CNNs trained on substantially larger datasets[10–13,15], underscoring its adaptability to a variety of camera characteristics and survey conditions. To assess the impact of the number of input examples on performance, we also evaluated Gemini using 3, 6 and 12 annotated image–text triplets instead of the full 15 MeerLICHT images. Our results show that, compared with the 15-triplet configuration, using 12 examples decreased accuracy by approximately 0.5 percentage points, using 6 examples led to a drop of around 1.4 percentage points, and using 3 examples resulted in a decrease of roughly 6.9 percentage points. This trend indicates that performance steadily improves with an increasing number of examples and plateaus around 15 examples, supporting the effectiveness of our chosen guide set size. To further assess the repeatability of our few-shot strategy, we reran the original 15-example guide set several times and also evaluated five new, non-overlapping sets with an identical class composition. The results of this analysis are presented in 'Six-month repeatability analysis with the updated Gemini 1.5-pro'.

Alongside each classification, Gemini provides a textual description detailing the observed features of the candidate. Although the accuracy of these classifications can be validated against known labels, evaluating the quality and coherence of the textual explanations requires alternative methods. To address this, we employ two complementary strategies: human evaluation by professional astronomers to ensure the outputs align with expert-level interpretation and a self-review mechanism within the model to identify and flag potential inconsistencies.

### Astronomers judging LLMs

To assess the clarity and fidelity of the explanations provided by Gemini, we conducted a human evaluation involving professional astronomers. A group of 12 astronomers participated in the survey using the Zooniverse platform (https://www.zooniverse.org/projects/fiorest/ai-on-trial-how-well-do-llms-classify-astronomical-images). Each participant reviewed a set of transients paired with the descriptions generated by Gemini. They rated the coherence of these explanations on a scale from 0 to 5, with the following definitions: 0 indicates a complete hallucination; 1 signifies that most of the explanation is incorrect; 2 means the explanation is more incorrect than correct; 3 denotes that the explanation is mostly correct with some errors; 4 corresponds to an almost entirely correct explanation; and 5 represents a perfectly coherent description. This scale reflects how well the textual descriptions captured the visual features in the new, reference and difference images.

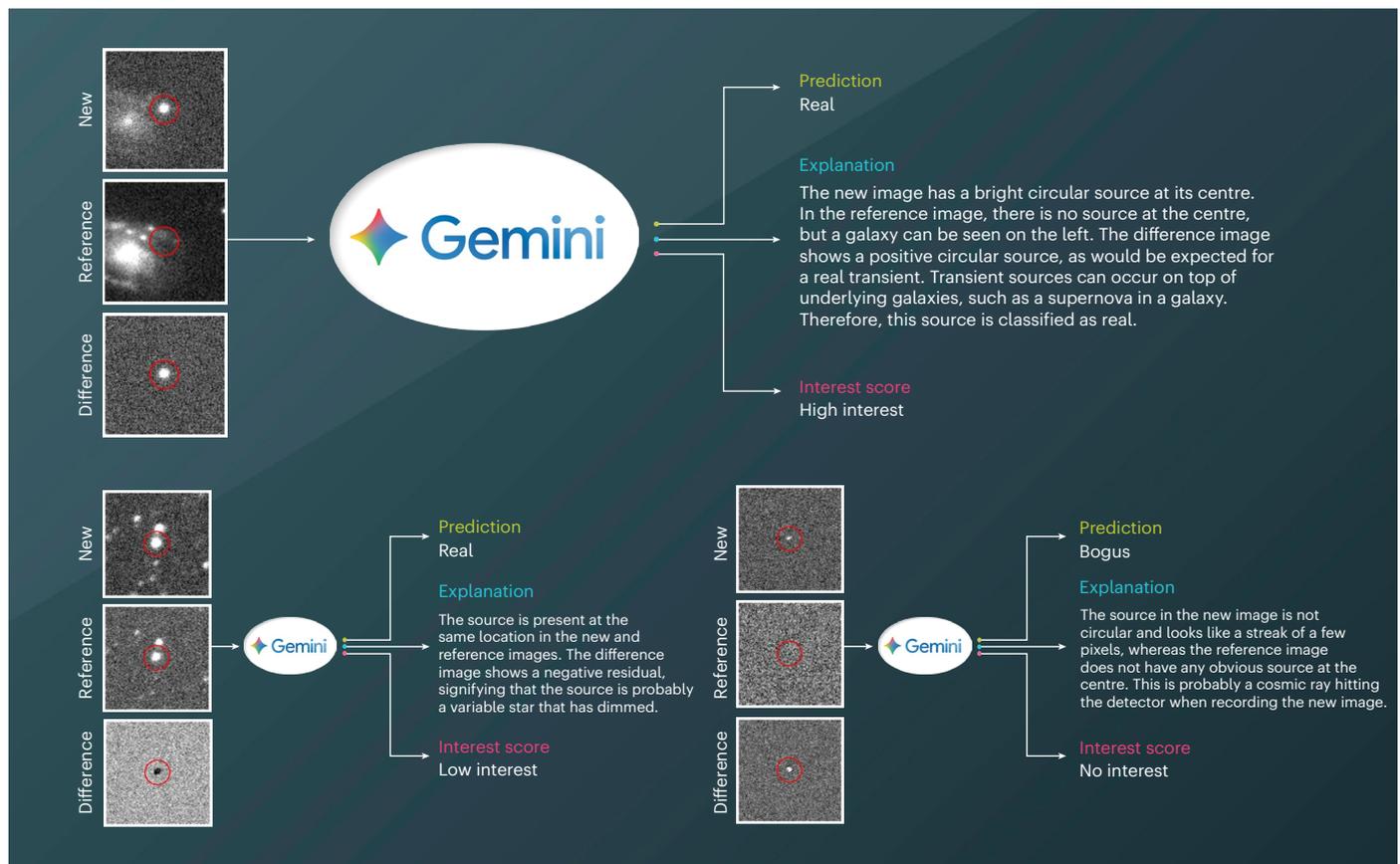

**Fig. 1 | Gemini provides human-readable transient classifications and follow-up priorities.** Each example consists of a new, reference and difference image for a candidate transient, followed by the Gemini classification, textual description and follow-up interest score. The examples shown in the figure are from the MeerLICHT dataset.

A total of 200 randomly selected MeerLICHT transients were evaluated by our panel of astronomers, with the following results. The mean coherence score exceeded 4, indicating that, on average, the explanations given by the LLM were perceived as highly accurate and meaningful. The distribution of scores was heavily concentrated around 4 and 5, showing that most descriptions closely aligned with the observed features. Notably, for approximately 120 out of the 200 images, there was complete agreement among the evaluators. For the remaining cases, the median standard deviation of the coherence scores was only 0.6, indicating that discrepancies between raters were minimal. For the interest score assigned by the model, the LLM demonstrated near-perfect self-consistency. Its stated interest level almost always matched the logic of its own explanations, with only one case where evaluators disagreed—one marking it as consistent and another as inconsistent—and no direct contradictions.

Figure 3 illustrates these findings. The left-hand panel shows that nearly all transients achieve high coherence scores, whereas the right-hand panel confirms that correct classifications by the model are closely associated with higher coherence ratings. These results indicate that Gemini not only makes accurate classifications but also produces explanations that are trusted and valued by human experts, facilitating greater confidence and transparency in automated astronomical data analysis workflows.

### LLMs judging LLMs
Beyond human evaluation, we investigated the capacity of Gemini to judge its own output using the same 0–5 coherence scale and interest scoring criteria. Like the astronomers, Gemini was set up to assign a coherence score to its explanations and verify whether its assigned interest level was logically consistent. First, Gemini produced an initial classification and a textual explanation for each MeerLICHT transient. Then it assigned a coherence score to this explanation and checked whether the interest assessment matched the narrative. Cases with low coherence scores were marked as potentially problematic.

Figure 4 shows that low coherence scores strongly correlate with misclassifications. In other words, Gemini effectively indicates which candidates it has likely gotten wrong. Even starting with just 15 input examples, this built-in self-awareness enables a highly targeted refinement process: by reviewing the low-coherence cases, adding a small number of these challenging examples to the input set and then rerunning the analysis, we can substantially improve the performance of the model.

Following this iterative strategy, we raised the initial accuracy for MeerLICHT from about 93.4% to approximately 96.7%. In essence, Gemini not only classifies transients but also identifies its own points of uncertainty, guiding a swift and efficient cycle of focused data augmentation and performance enhancement, even when the initial labelled data are limited.

### Discussion
Our study demonstrates that LLMs like Gemini offer a transformative approach to astronomical transient classification. Gemini achieves high accuracy with minimal labelled data and provides detailed, human-readable explanations alongside its classifications. In contrast to traditional methods—including CNNs and emerging vision transformer techniques[29,30] that rely on auxiliary interpretability tools such as Grad-CAM—Gemini directly produces textual explanations. Although this integrated interpretability comes with increased computational demands, it offers a valuable complement in scientific contexts by providing explanations that are accessible and verifiable by human experts.

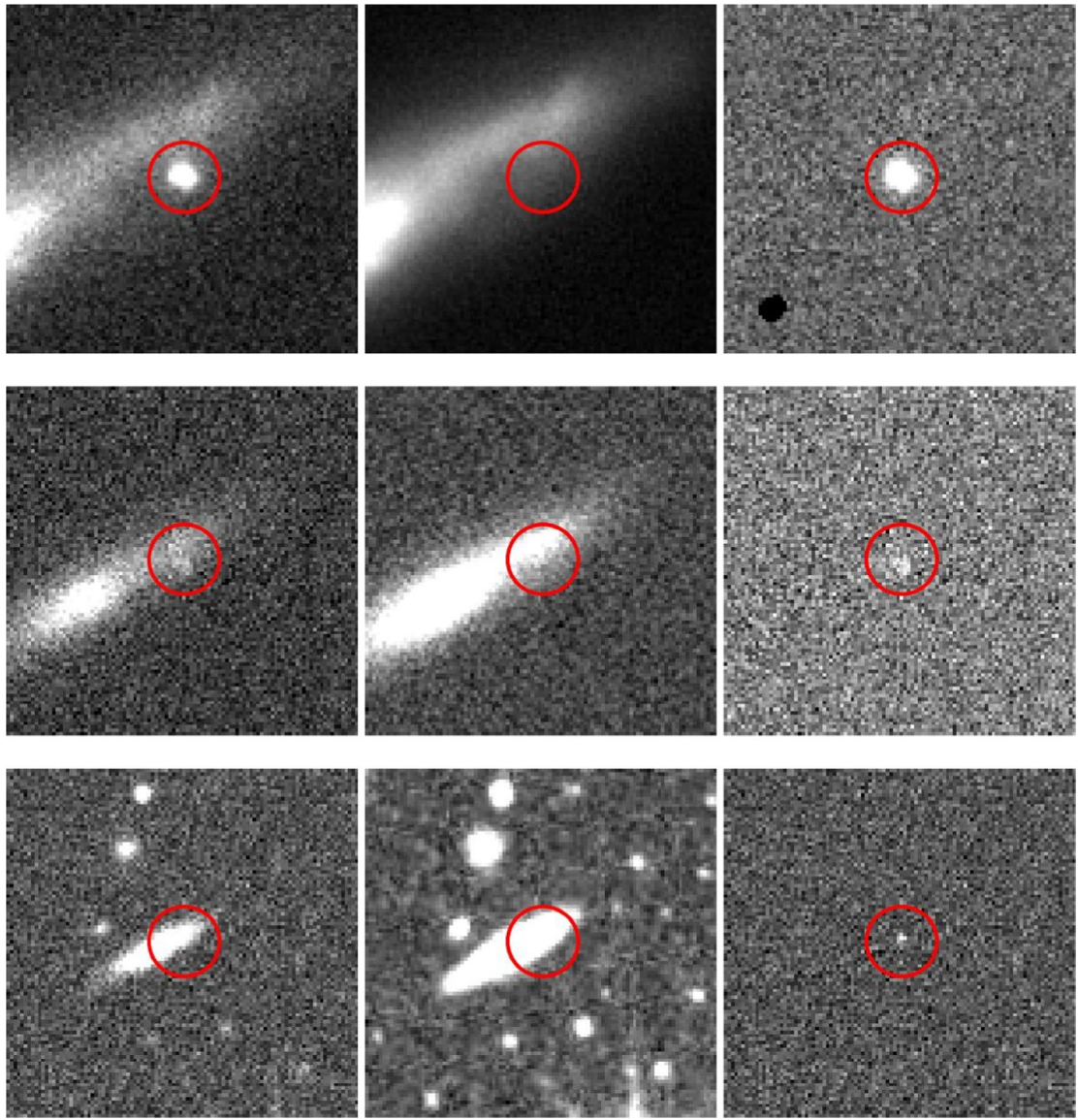

**Fig. 2 | Gemini operates across surveys with diverse pixel scales and resolutions.** The same transient observed in three different surveys, with rows corresponding to Pan-STARRS (top row), MeerLICHT (middle row) and ATLAS (bottom row). Each row includes, from left to right, a new image, a reference image and a difference image. Red circles mark the expected position of the transient candidate at the centre of each stamp. The image stamps are all the same size in pixels (100 × 100) but differ in angular sky coverage due to survey-specific pixel scales: Pan-STARRS (0.25" per pixel), MeerLICHT (0.56" per pixel) and ATLAS (1.8" per pixel).

One of the most striking aspects of our findings is that Gemini required only a written set of instructions and 15 example triplets per dataset to achieve high classification accuracy (a list of the instructions can be found in 'Gemini model and prompts'). Unlike traditional machine learning models, which require extensive labelled datasets and frequent retraining to adapt to new data, the few-shot learning capabilities of Gemini allow it to generalize effectively with minimal input, also across different surveys, for which no implementation exists at all yet. This efficiency reduces the time and resources typically required for model development and maintenance. The ability to use minimal input examples is especially advantageous in the context of transient astronomy, where new types of event can emerge and the characteristics of known transients can evolve over time. The adaptability of Gemini means that it can be quickly updated with new examples or instructions, thus ensuring that the classification system remains current without the need for exhaustive retraining. This adaptability is not exclusive to LLMs; traditional CNNs can be optimized for low-data scenarios using strategies such as domain adaptation[31] or active learning[16], yet employing LLMs offers a considerably simpler approach.

Traditional machine learning methods often rely on complex latent space representations to summarize data[32,33]. Although effective in many contexts, these latent spaces can be opaque, making it difficult for astronomers to interpret what the model has learned or to interact with the data in meaningful ways. By contrast, Gemini provides an accessible and understandable summary of each event in the form of descriptive sentences. This approach aligns with the way that astronomers traditionally communicate and reason about events. By generating textual descriptions that highlight observable

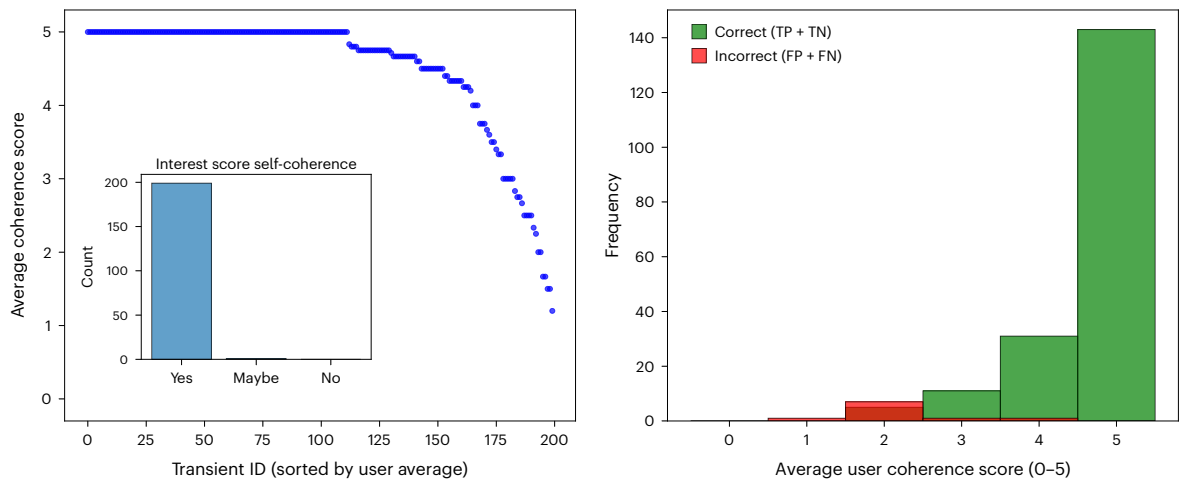

**Fig. 3 | Expert ratings show that the explanations provided by Gemini are generally consistent and coherent.** Left: average coherence scores from 12 astronomers for 200 MeerLICHT transients, sorted by mean score (blue markers). Most examples received high coherence values (4–5), indicating close alignment with user expectations. Inset: the bar chart shows the consistency between the interest score assigned by the model and its own explanation, with nearly all cases marked as self-consistent (yes). Right: histogram of average user coherence scores, split by the correctness of the classification made by Gemini. Correctly classified examples (TPs and TNs, green) tend to have higher coherence scores than incorrect ones (FPs and FNs, red), indicating a link between accuracy and explanation quality.

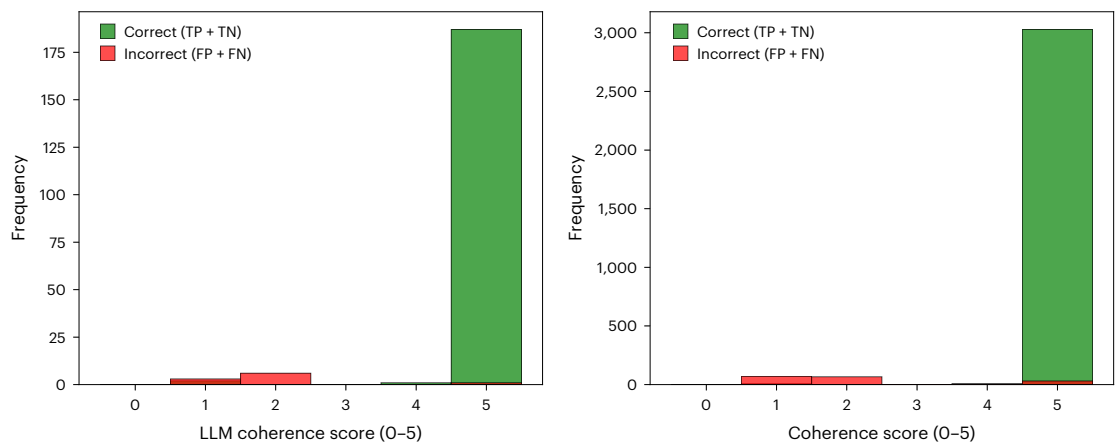

**Fig. 4 | Coherence scores self-assigned by Gemini correlate with classification correctness.** Left: histogram of self-assigned coherence scores for the same 200 MeerLICHT transients evaluated by professional astronomers in Fig. 3, split by whether the classification was correct (TPs and TNs, green) or incorrect (FPs and FNs, red). Misclassified cases tend to receive lower coherence scores. Right: same histogram extended to the full MeerLICHT dataset, showing a similar trend across thousands of examples. This pattern indicates that the coherence estimates given by Gemini can be used to flag potentially uncertain classifications.

features—such as the shape, brightness and variability of sources—Gemini allows for straightforward interpretation and verification of its classifications. Moreover, the textual outputs enable efficient data-querying and retrieval. A second LLM could be employed to analyse the outputs and transient explanations generated by our LLM from one night of observations. This second LLM would interactively sift through the descriptions to identify specific types of objects of interest. For example, astronomers could query the system for transients likely associated with galaxies, isolated orphan events or patterns indicative of rare phenomena. This interactive approach would allow researchers to efficiently pinpoint the most relevant candidates, thus generating actionable insights from large observational datasets without the need to visualize or process individual images.

However, the practical deployment of LLMs for real-time astronomical surveys demands careful consideration of computational performance. Traditional CNN-based pipelines can process individual images in milliseconds on GPUs, whereas large LLMs typically require several seconds per query due to high token-generation latency. This difference becomes critical when considering the scale of upcoming surveys such as the Vera Rubin Observatory, which is expected to generate of the order of 10 million alerts per night. Running a large LLM through a commercial API under these conditions is, at present, impractical, not only due to the extended inference times but also because of prohibitive costs that could amount to thousands of dollars per night. This is, however, a fast-evolving field. Promising avenues for optimizing performance include fine-tuning smaller open-source LLMs with one to two billion parameters or applying model quantization techniques to enable faster inference on local hardware[34–36]. Another alternative is a hybrid approach, where fast CNNs perform an initial screening and the LLM is engaged selectively for detailed interpretation of interesting or ambiguous cases.

Beyond transient image classification, the approach demonstrated by Gemini has potential applications across various domains in astronomy. For instance, it could be used in the classification of galaxies, the identification of exoplanet transits or analyses of gravitational lensing events. Any task that involves interpreting complex image data could benefit from the combination of few-shot learning and natural language descriptions. Furthermore, the methodology can be extended to multimodal data analysis by integrating spectral information and time-series data. Incorporating such diverse modalities would represent a substantial step forward, transitioning from just detecting sources to enabling a comprehensive scientific analysis and classification of the type of event.

Although the results are promising, several challenges remain. The reliance on LLMs raises questions about model biases and the need for careful prompt engineering to ensure accurate and reliable outputs[37,38]. Recent literature has begun developing tools to identify potential biases at the data, model and output levels, including sensitivity to prompt formats[39–41]. Although no study has yet specifically investigated this issue in astronomy, it is plausible that if the few-shot examples used to guide Gemini do not adequately represent the full diversity of transient phenomena—such as variations in observational conditions or a comprehensive mix of astronomical sources—the classifier may inadvertently favour certain classes over others. Furthermore, it is important to note that effective prompts are often unique to specific LLMs, meaning that a prompt that works well with one model may not necessarily generalize to another. In this regard, recent studies[42–44] have shown that variations in prompt wording, example order, formatting (plain text, bullet points or JSON) and chain of thought prompting can produce notable differences in task performance and interpretability.

Future work could explore how to improve results through supervised fine-tuning or reinforcement learning with human feedback, which would blend the power of LLMs with astronomer expertise to create adaptive, high-performing models[45–48]. Additionally, creating repositories of diverse examples and developing task-specific instructions for particular astronomical applications could help advance the broader use of LLMs in astronomy. Investigating hybrid approaches that combine LLMs with other machine learning models may also enhance performance by leveraging the complementary strengths of different methodologies. Overall, our findings highlight the potential of LLMs to integrate transparent, human-like reasoning into transient classification, setting the stage for further domain-specific refinements and more interpretable astronomical analyses.

## Methods
### Gemini model and prompts
This section provides details on the set-up and configuration of the Gemini LLM, including the structured prompts, instructions and task definitions used to guide its classifications. We used Google's LLM, gemini-1.5-pro-002, accessed via the Google Cloud Platform.

Gemini was guided by structured prompts designed to emulate the decision-making process of an experienced astronomer. This prompt engineering ensured that the responses of the model aligned with the domain-specific knowledge required for astronomical classification. Note that current LLMs generate outputs that mimic expert reasoning solely by identifying patterns in the input data based on their training, without any genuine understanding, awareness or human-like cognition. The key aspects of our prompt design included:

- Persona definition: The model was instructed to adopt the role of an expert astrophysicist and provide responses with domain-specific terminology and insights.
- Explicit instructions: Detailed guidelines were provided for identifying critical features of real and bogus transients, focusing on observable attributes such as shape, brightness and variability.
- Task clarification: Each prompt clearly outlined the task: to classify the transient as real or bogus, to describe the features observed in the images that led to this classification and, where applicable, to assign an interest score.
- Few-shot learning examples: A limited set of real and bogus examples (15 per dataset) was provided, as the model could generalize effectively with minimal training data.

Below is the complete set of prompts used for the MeerLICHT application. The full collection of instructions, prompts and examples is available in the associated GitHub repository (https://github.com/turanbulmus/spacehack).

| Persona definition |
|---|
| You are an experienced astrophysicist, and your task is to classify astronomical transients into Real or Bogus based on a given set of 3 images. You have seen thousands of astronomical images during your lifetime and you are very good at making this classification by looking at the images and following the instructions. |

| Instructions for Real/Bogus Classification |
|---|
| 1. Purpose |
| Help vet astronomical data for the Real/Bogus classification. The goal is for you to use your expertise to distinguish between real and bogus sources. |
| 2. Information Provided |
| You will be shown three astronomical image cutouts: |
| a) New Image: The newest image centred at the location of the suspected transient source. |
| b) Reference Image: A reference image from the same telescope of the same part of the sky to be used for comparison. It shows if the source was already there in the past or not. |
| c) Difference Image: The residual image after the new and reference images are subtracted. Real sources should appear in this cutout as circular objects with only positive (white pixels) or only negative (black pixels) flux. |
| 3. Criteria for Classification |
| Real Source: |
| - Shape: Circular shape at the centre of the cutout with a visual extent of ~5-10 pixels, varying with focus conditions. |
| - Brightness: Positive flux (white pixels) in either the new or reference image. Positive or negative flux in the Difference image. |
| - Variability: The source at the centre can fade or brighten between the new and reference images, appearing as positive or negative in the Difference image. |
| - Presence: The source may (dis)appear between the new and reference images. A source may also appear on top of an underlying source (for example, supernova on a galaxy). |
| Bogus Source: |
| - Shape: Non-circular shape (for example, elongated). This includes irregular shapes, positive or negative, like streaks or lines caused by cosmic-rays, diffraction spikes and cross-talk. |
| - Brightness: Negative flux (black pixels) at the centre of the cutout in either the new or reference image. The source at the centre can never be negative in the New or Reference image, only in the Difference. |
| - Misalignment: If the source in the New and Reference images is misaligned, it will show a Yin-Yang pattern (both white and black) in the Difference image. |
| 4. Additional Guidance |
| Contextual Information: Focus on the source at the centre of the cutouts inside the red circle, but consider nearby sources to diagnose potential problems. |
| Examples: Refer to provided visual examples of real and bogus sources to aid in identification. |
| Judgment Criteria: For ambiguous cases or borderline scenarios, consider the overall context and consistency with known characteristics of real and bogus sources. |

| Method definition |
|---|
| 1. Focus on the Red Circle: Start by examining the source located at the centre of the cutout and inside the red circle. The images are prepared so that the source of interest is clearly marked for you to analyze. |
| 2. Analyze Each Image Individually: |
| - New Image: Check for the presence, shape, and brightness of the source in the new image. |
| - Reference Image: Compare the source's properties in the reference image to those in the new image. |
| - Difference Image: Observe the residuals that result from subtracting the reference image from the new image. Look for patterns (circular, positive/negative flux) that match characteristics of Real or Bogus sources. |
| 3. Evaluate Features: |
| - Examine the shape, brightness, and other relevant features (for example, artifacts, misalignments) of the source in each image. |
| - Determine if these features are consistent with a Real or Bogus classification based on the criteria provided in the instructions. |
| 4. Consider Relationships Between Images: |
| - Compare the new, reference, and difference images to understand any changes in the source over time. |
| - Look for discrepancies or confirmations that might support or contradict a particular classification. |
| 5. Employ a Chain-of-Thought Reasoning: |
| - Clearly outline each observation you make and explain how it contributes to your decision-making process. |
| - If you find any contradictions or ambiguous features, acknowledge them and provide reasoning for your final decision. |
| 6. Assign an Interest Score: |
| - After determining if the source is Real or Bogus, assign an appropriate interest score: |
| - 'No interest' for Bogus sources. |
| - 'Low interest' for variable transients. |
| - 'High interest' for explosive transients. |
| 7. Prepare the Final Output in JSON Format: |
| - Format your response as a JSON object containing: |
| - The classification ('Real' or 'Bogus'). |
| - An explanation detailing your thought process and observations. |
| - The assigned interest score. |
| 8. Example Output: |
| - Refer to the provided examples to see the expected format and detail level of your response. |

**Few-shot learning examples from the MeerLICHT dataset**
Below are the 15 few-shot learning examples used to guide the classifications made by Gemini for the MeerLICHT dataset. These are presented in Supplementary Fig. 1 (bogus examples) and Supplementary Fig. 2 (real examples).

**Descriptions of bogus examples.**
*Example 1.*
   **Class:** Bogus
   **Interest score:** No interest
   **Explanation:** In the new image, a diffraction spike is observed near the centre. The reference image also shows a diffraction spike at the same location. In the difference image, a negative residual of the bright diffraction spike from the reference image is clearly visible. The consistent presence of diffraction spikes in all three images, without a clear circular source, confirms that this is a bogus source.

*Example 2.*
   **Class:** Bogus
   **Interest score:** No interest
   **Explanation:** In the new image, a negative elongated artefact is present at the centre. The reference image does not show any source at the same location. In the difference image, the same negative artefact appears, which results from the negative clump of pixels in the new image. As a real source cannot be negative in the new image, this is classified as a bogus source.

*Example 3.*
   **Class:** Bogus
   **Interest score:** No interest
   **Explanation:** In the new image, the source appears as a streak of several bright pixels and is not circular. The reference image shows no source at the same location. The difference image shows the same streak of pixels as in the new image. The sharp, streak-like appearance in the new image indicates that this is most probably a cosmic ray rather than a real source.

*Example 4.*
   **Class:** Bogus
   **Interest score:** No interest
   **Explanation:** The new image does not have any source at the centre of the cut-out. The reference image shows a source appearing as a streak of a few bright pixels, which is not circular. The difference image shows the negative residual of the same streak present in the reference image. This is too sharp to be a real source and is probably a cosmic ray that was not flagged during the creation of the reference image.

*Example 5.*
   **Class:** Bogus
   **Interest score:** No interest
   **Explanation:** No source is present in the new image. In the reference image, a source appears as a negative circular object. The difference image presents a faint positive residual of the source in the reference image. As a source cannot be negative in the reference image, this is not a real source.

*Example 6.*
   **Class:** Bogus
   **Interest score:** No interest
   **Explanation:** The new image does not have any source at the centre of the cut-out. In the reference image, the source appears very elongated. The difference shows the same negative elongated source, supporting the conclusion that it a bogus source.

*Example 7.*
   **Class:** Bogus
   **Interest score:** No interest
   **Explanation:** In the new image, a small, elongated source is visible and surrounded by several other sources. The reference image shows no source at the same location, but it does show all the other sources. In the difference image, the residual is positive but its elongation confirms that this is a bogus source.

*Example 8.*
   **Class:** Bogus
   **Interest score:** No interest
   **Explanation:** The new image shows a diffuse source at the centre, aligned with a 45° diffraction spike from a bright source at the corner of the cut-out. The reference image also shows a diffraction spike and a similar blob. The difference image displays a positive blob, indicating it is an artefact caused by the diffraction spike, which can produce blobs or irregular shapes.

*Example 9.*
   **Class:** Bogus
   **Interest score:** No interest
   **Explanation:** The new image shows no source at the centre. The reference image shows a faint positive trail cutting diagonally across the image with a circular source at the centre, which was probably caused by a blinking object like an aeroplane or satellite. The difference image displays both the trail and a negative blob, confirming that the source is a non-astronomical artefact.

**Descriptions of real examples.**

*Example 10.*
   **Class:** Real
   **Interest score:** Low interest
   **Explanation:** The new image shows a source at the centre. The reference image also shows the same source in the same location. The difference image has a positive residual, indicating that the source has brightened. This pattern indicates that the source is a real variable star.

*Example 11.*
   **Class:** Real
   **Interest score:** Low interest
   **Explanation:** The new image shows a source at the centre. The reference image also shows the same source in the same location. The difference image has a negative residual, indicating that the source has dimmed. This pattern indicates that the source is a real variable star.

*Example 12.*
   **Class:** Real
   **Interest score:** High interest
   **Explanation:** The new image shows no source at the centre. The reference image shows a circular source in the same location. The difference image displays a negative circular residual, consistent with a transient that has disappeared.

*Example 13.*
   **Class:** Real
   **Interest score:** High interest
   **Explanation:** The new image shows a bright circular source at the centre. The reference image shows no source in the same location. The difference image displays a positive circular residual, indicating a real explosive transient.

*Example 14.*
   **Class:** Real
   **Interest score:** Low interest
   **Explanation:** The new image shows a source at the centre. The reference image also shows the same source in the same location. The difference image displays a positive residual, indicating that the source has brightened. A cosmic ray artefact is visible to the left, but the central source is unaffected and remains a valid transient.

*Example 15.*
   **Class:** Real
   **Interest score:** High interest
   **Explanation:** The new image shows a source at the centre and superimposed on a diffuse galaxy. The reference image displays the galaxy but no source at the same location. The difference image reveals a faint, positive circular feature, consistent with a supernova emerging within the galaxy.

## Six-month repeatability analysis with the updated Gemini 1.5-pro

Understanding the repeatability of results is critical when evaluating few-shot prompting strategies, particularly in the context of rapidly evolving LLMs. To assess the robustness and reproducibility of our findings, we conducted a full reanalysis of the MeerLICHT experiment approximately 6 months after the results reported in 'Results' and 'Discussion', using the same gemini-1.5-pro end point, now updated with newer weights and potential decoding refinements.

We reconstructed 5 new, non-overlapping sets of 15 triplets from the original MeerLICHT dataset, which were designed to match the object class balance of the original exemplar group. Importantly, we held constant all other variables: prompt structure, decoding parameters and evaluation code. Each of these 5 sets was evaluated across 5 independent Gemini runs, yielding a total of 25 inference batches.

The analysis reveals consistently low levels of variability. Within each set, classification metrics varied only minimally ($\sigma_{Acc} = 2.99 \times 10^{-4}$, $\sigma_{Prec} = 5.90 \times 10^{-4}$ and $\sigma_{Rec} = 2.32 \times 10^{-4}$, for accuracy, precision and recall, respectively). Even across different sets, between-group standard deviations remained modest ($2.34 \times 10^{-3}$ for accuracy, $6.63 \times 10^{-3}$ for precision and $2.64 \times 10^{-3}$ for recall), demonstrating that the method is robust to both stochastic sampling and exemplar composition.

Because this evaluation was conducted half a year after the original study, the underlying model had undergone updates. Although our primary aim here was a repeatability assessment, note that the updated model exhibited improved average performance: accuracy increased from 0.934 to 0.962 (+2.8 percentage points), precision from 0.877 to 0.929 (+5.2 percentage points) and recall from 0.987 to 0.992 (+0.5 percentage points). These gains, shown in Supplementary Fig. 3, are a secondary yet important observation: although previous results may not be exactly reproducible with commercial LLMs, the broader performance trend indicates steady improvement even within nominally stable model names.

Overall, this repeatability analysis confirms that the few-shot classification pipeline is stable across runs, consistent across different example choices and resilient to moderate underlying model updates. However, it also highlights a pragmatic concern for future work: in research pipelines that depend on commercial LLMs, periodic revalidation should be expected and operationally planned for, especially as models evolve behind version-stable end points.

## Data availability

The data used in this paper, including the annotated examples, transient images and the Zooniverse project results, are available via Zenodo at https://doi.org/10.5281/zenodo.14714279 (ref. 49).

## Code availability

All code used for generating the prompts, interacting with the Gemini models and analysing the results is available via GitHub at https://github.com/turanbulmus/spacehack/. This repository allows for full replication of the findings of the study and facilitates further exploration of LLMs in astronomical transient classification.

## Acknowledgements

F.S. and T.B. contributed equally to this work. F.S. acknowledges support from the Royal Society Newton International Fellowship (NIFR1241769). S.J.S. acknowledges funding from the STFC (grant nos. ST/Y001605/1, ST/X006506/1 and ST/T000198/1), a Royal Society Research Professorship and the Hintze Charitable Foundation. P.J.G. is supported by the NRF (SARChI grant no. 111692). The MeerLICHT telescope was designed, built and operated by a consortium consisting of Radboud University, the University of Cape Town, the South African Astronomical Observatory, and the Universities of Oxford, Manchester and Amsterdam.

## Author contributions

F.S. wrote the paper and developed the core analysis code. T.B. contributed to code development and served as the Gemini LLM expert. S.B. initiated project collaboration and coordinated the early development. P.J.G. provided expertise on MeerLICHT data and contributed to reviewing the paper. P.V. supported the data processing and retrieval for MeerLICHT observations. K.W.S. processed and retrieved the ATLAS and Pan-STARRS data. S.J.S. advised on the ATLAS and Pan-STARRS integration and reviewed the paper.

## Competing interests

The authors declare no competing interests.

## Additional information

**Supplementary information** The online version contains supplementary material available at https://doi.org/10.1038/s41550-025-02670-z.

**Correspondence and requests for materials** should be addressed to Fiorenzo Stoppa.

**Peer review information** *Nature Astronomy* thanks Zhonghao Chen, Samuel Farrens and the other, anonymous, reviewer(s) for their contribution to the peer review of this work.

**Reprints and permissions information** is available at www.nature.com/reprints.

**Publisher's note** Springer Nature remains neutral with regard to jurisdictional claims in published maps and institutional affiliations.